\documentclass[useAMS,usenatbib]{raa}
\usepackage{amssymb}
\usepackage{amsmath}
\usepackage{graphicx}
\usepackage{lscape}
\usepackage{color}
\usepackage{subfigure}

\begin{document}

 \title
{What is the right way to quench star formation in semi-analytic model of galaxy formation?}

 \author
{Yu Luo\inst{1,2}, Xi Kang\inst{1} }

\institute {Purple Mountain Observatory, the Partner Group of MPI f\"ur Astronomie, 2 West Beijing Road, Nanjing 210008, China; {\it luoyu@pmo.ac.cn,kangxi@pmo.ac.cn}\\
\and 
Graduate School, University of the Chinese Academy of Science, 19A, Yuquan Road, Beijing 100049, China\\
}

\abstract{Semi-analytic models of galaxy  formation are powerful tools
  to  study  the evolution  of  galaxy  population in  a  cosmological
  context. However,  most models  over-predict the number  of low-mass
  galaxies at high  redshifts and the color of model  galaxies are not
  right in the sense that low-mass  satellite galaxies are too red and
  centrals are too blue. The recent version of the L-Galaxies model by
  Henriques  et  al.(H15)  is  a  step  forward to  solve  these
  problems by reproducing  the evolution of stellar  mass function and
  the overall fraction of red galaxies.   In this paper we compare the
  two model predictions  of L-Galaxies (the other is Guo  et al. ,
  G13) to the SDSS data in detail. We find that in the H15 model the
  red fraction  of central galaxies  now agrees  with the data  due to
  their implementation of strong AGN feedback, but the stellar mass of
  centrals in massive haloes is now  slightly lower than the data. For
  satellite  galaxies, the  red  fraction of  low-mass galaxies  ($\log
  M_{*}/M_{\odot} < 10$)  also agrees with the data,  but the color
  of  massive  satellites ($10  <  \log  M_{*}/M_{\odot} <  11$)  is
  slightly bluer.   The correct color  of centrals and bluer  color of
  massive satellites indicate that the quenching in massive satellites
  are not  strong enough.  We also  find that there  are too  much red
  spirals  and less  bulge-dominated galaxies  in both  H15 and  G13
  models. Our results suggest that additional mechanisms, such as more
  minor merger  or disk instability,  are needed to  slightly increase
  the stellar  mass of central  galaxy in massive galaxies,  mainly in
  the bulge  component, and  the bulge dominated  galaxy will 
  be quenched or then be quenched by any other mechanisms.
}

   \authorrunning{Yu Luo \& Xi Kang}            
   \titlerunning{Galaxy quenching in SAMs}  
\maketitle

\section{Introduction}

The standard cold dark matter  model is now very successful to explain
the structure formation in the  universe from the very early beginning
up to the present (see the  review by Frenk \& White 2012). The nature
of  dark matter  and dark  energy  is the  main science  goal of  next
generation sky  surveys, such as EUCLID  and LSST. On  the other hand,
the wealth data of galaxy from  the ongoing and future surveys is very
useful for the study of  galaxy formation and evolution. Although many
efforts have been made to modelling of galaxy formation, it is fair to
say that there is no any model which can simultaneously well reproduce
the various properties of galaxies from the local surveys, for example
the  Sloan Digital  Sky Survey.  In that  sense, our  understanding of
galaxy  formation lags behind  our knowledges  about the  evolution of
dark  structure  which is  accessible  using  large N-body  simulations
(e.g., Springel et al. 2005) .

Hydro-dynamical  simulation  and  semi-analytic  models  are  powerful
physical  tools   to  study  galaxy  formation  and   evolution  in  a
cosmological context. The state-of-art hydro-dynamical simulations are
now being  able to reproduce  many properties of galaxies  in typical
cosmological volumes  (e.g., Vogelsberger et  al.  2014; Schaye  et al.
2015),  but   due  to   the  expensive  cost   of  running   huge  and
high-resolution hydro-dynamical  simulations, it is  difficult to test
how galaxy  properties correspond to  adopted physical models
of star formation and feedback.  Unlike hydro-dynamical simulations, SAMs are based on
N-body simulations, and they take simple phenomenological descriptions
of  the   physical  precesses about galaxy  formation,   such  as
cosmic reionization, hot gas cooling and cold gas infall, star formation and
metal production,  SN feedback, gas stripping and  tidal disruption of
satellites, galaxy mergers, bulge formation, black hole growth and AGN
feedback. SAMs  are developing  quickly in the  past decades  and most
basic features of galaxy formation are already included  in the earlier
versions  of SAMs  (e.g. White  \& Frenk  1991; Somerville  \& Primack
1999; Kang et  al. 2005; Bower et al. 2006;  Somerville et al.  2008).
The  main  goal of  SAMs  is  to reproduce  as  more  as possible  the
statistical properties of galaxies seen in the data. What is more important
of SAM is not to make predictions of new observations, but to test our
understanding of physical model of galaxy formation in a fast and easy
way.

SAMs have achieved great progress in recent years.  They can reproduce
the observed local stellar mass function down to the low mass end (Guo
et  al.  2011, Kang  et  al.  2012), cold  gas  mass  function (Fu  et
al. 2013), and some  scaling relations like the Tully-Fisher relation
(Guo et al. 2011,  hereafter G11), Faber-Jackson relation (Tonini et
al. 2016).  However, most  (if not all)  models can not  reproduce the
color  distribution  of galaxies.   The  color  of low-mass  satellite
galaxies  are too  red (e.g.,  Weinmann et  al. 2006),  the  color of
central galaxies  are too blue (Kang  et al. 2006).  These problem are
also persist in the model of  Guo et al. (2013, hereafter G13) which
have  included  more  advanced  treatments  of gas  cooling  and  star
formation in satellite galaxies.  The color discrepancy indicates that
quenching  of low  mass galaxies  is  too strong,  while quenching  of
massive galaxies is  not enough.  So the quenching  mechanism might be
different  in  low mass  and  high  mass  galaxies: the  low-mass  red
galaxies in  models are mainly satellites which are often  
related to environment  quenching, like
ram-pressure and tidal stripping.   However, Luo et al. (2016) pointed
it is not purely of environmental origin. The blue massive galaxies in
models is related to insufficient feedback.

The recent version  of the L-Galaxies model, Henriques  et al. (2015),
has  achieved  progress  to   solve  the  color  distribution  in  the
model. They use  Markov Chain Monte Carlo (MCMC)  method to search the
parameter  space to  simultaneously fit  the stellar  mass  function at
different redshifts and the overall red fraction of galaxies. Although
their   success  comes  at   no  surprise,   it  indicates   that  our
understanding of galaxy  formation is on the right  way. For detail of
how they  succeed in  reproducing the two  observations, we  refer the
readers to  their paper.  However, it  is still not  clear of  the H15
model  can also  reproduce  the color  distribution  of satellite  and
central galaxies respectively since  the quenching mechanism might be
different for  satellite and central  galaxies. In this paper  we will
compare the model predictions of G13  and H15 on the red fraction of
satellite  and  central  galaxies,  which  is the  main  goal  of  our
paper.  On  the other  hand,  it is  well  known  that star  formation
quenching is close related  to galaxy morphology that most red/passive
galaxies are  bulge dominated (e.g.,  Kauffmann et al. 2004),  we also
use the data to check the morphology mix in the model.

This paper is organized as follows. In Section 2, we briefly summarize
the L-Galaxies models  and describe the main modifications  in H15. In
Section 3, we  analyze the galaxy quenching in  H15, comparing them to
the  G13  model  and  the  SDSS data,  such  as:  quenching  fraction,
morphology of central galaxies, conditional stellar mass functions and
stellar  mass-halo  mass relation.  In  Section  4,  we summarize  our
results and discuss the possible improvement.

\section{The L-Galaxies semi-analytic models}
In this section, we briefly introduce the L-Galaxies models and the 
simulation they used, then describe the main changes of physics recipes 
between H15 and G13.
\subsection {The models and simulations}
The  L-Galaxies is  one  of the  most  successful semi-analytic  galaxy
formation  model.   It  is  continuously  developed   by  Kauffmann  et
al.(1999); Springel  et al.(2001);  Croton et  al.(2006); De  Lucia \&
Blaizot (2007) in last two  decades. Recently, the L-Galaxies model of
G11 and G13 had been improved comprehensively which can well reproduce
observed galaxy populations from dwarf spheroidals to cD galaxies.  Fu
et al. (2010 \& 2013) have developed  a model in which the cold gas is
partitioned into atomic and molecular components. Yates et al. (2014)
adopted a  more realistic  model for  chemical evolution.  After that,
Henriques  et al.(2015)  published  the latest  trunk  version of  the
L-Galaxies,  which  is claimed  to  match  the observed  evolution  of
stellar mass  function, galaxy  colors, star-formation rates,  in the
\emph{Planck} cosmology. In this work we  will use the data produced from the
G13 and H15 models.

Both G13  and H15  models are implemented  on two  N-body simulations:
Millennium  (here after  MS, Springel  et al.  2005) and  Millennium-II
(here after MS-II,  Boylan-Kolchin et al. 2009).  Both simulations use
the  cosmological  parameters  from   the  first-year  WMAP  data  and
$2160^{3}$  particles. The  simulation box  of MS  is $500Mpc/h$,  and
MS-II  has box  size of  $100Mpc/h$, so  it is  125 times  higher mass
resolution than MS.  Using the technique developed by  Angulo \& White
(2010) and  Angulo \&  Hilbert (2014),  G13 rescaled  the cosmological
parameters    from    WMPA1     to    WMAP7:    $\Omega_\Lambda=0.728,
~\Omega_m=0.272,~\Omega_{\rm{baryon}}=0.045,~\sigma_8=0.807$       and
$h=0.704$,    while     H15    rescaled    to     \emph{Planck}    cosmology:
$\Omega_\Lambda=0.685,
~\Omega_m=0.315,~\Omega_{\rm{baryon}}=0.0487,~\sigma_8=0.829$      and
$h=0.673$.

H15 applied the MCMC procedures to find best parameter by fitting to 
a series of fiducial observational data: the evolution of stellar mass 
function and the passive galaxy fraction as a function of stellar mass 
from $z=0$ to $z=3$. Both  galaxy catalogues from H15 
and G13 are publicly available on Millennium data base. 
\footnote {http://gavo.mpa-garching.mpg.de/Millennium}

\subsection{The main modifications in H15 model}

Compared to G13,  H15 modified a few treatments  of baryonic processes
to best  fit the observed  evolution of star-formation  rates, colours
and  stellar   mass  up   to  $z=3$.  For   more  details   about  the
modifications, please refer the supplementary material in H15. Here we
list the main modifications by H15 compared to G13 model:

\noindent(i)  delay the  reincorporation rate  of ejected  gas into  the
halo: changing the ejected gas reincorporation rate from $\dot{M}_{\rm
  ejec}  \propto -V_{\rm  vir}/t_{\rm dyn,h}$  to $\dot{M}_{\rm  ejec}
\propto   -M_{\rm  vir}$.   This  change   leads  to   a  slower   gas
reincorporation rate in low-mass  haloes and a quicker reincorporation
rate in massive haloes.

\noindent(ii)  decrease  the  threshold density  for  star  formation:
changing   the   gas   density  threshold   $\Sigma_{\rm   SF}$   from
$3.8\times10^9M_\odot/pc^{2}$  to   $2.6\times10^9M_\odot/pc^{2}$.  In
this way satellite galaxy with poor gas can still form stars.

\noindent(iii)  ignoring the  ram-pressure stripping  in less  massive
haloes: setting a  ram-pressure threshold $M_{r.p.}=1.2\times10^{14}M_{\odot}$.
So the  hot gas of satellites  in low-mass haloes is  not stripped and
available for continuous cooling and star formation.

\noindent(iv)  increasing   the  radio-mode  AGN  feedback:   the  new
efficiency  of  hot  gas  accretion   onto  the  central  black  hole,
$\kappa_{\rm  AGN}$, is  equivalent to  the one  in G13  divided by  a
factor of  $H(z)$. The larger accretion  efficiency in H15 leads  to a
quicker growth  of black hole  and a  stronger AGN feedback  (which is
related to  gas accretion  rate). So  the suppress  of gas  cooling in
mediate and massive galaxies is stronger in H15 model.

\section{Model predictions}

In this section, we compare the  model predictions of H15 and G13 to
the data  on the red fraction  of central and satellite  galaxies, the
fraction of galaxies  with different morphology, and  the stellar mass
to halo mass relation.

\begin{figure}
\centering
\includegraphics[scale=0.7]{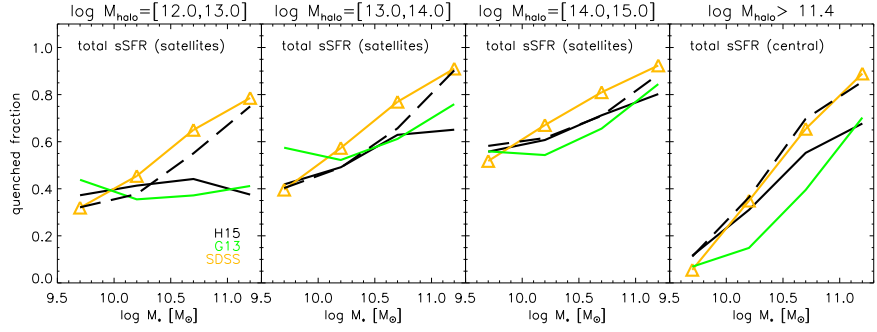}
\caption{The quenched  fraction of galaxies  as a function  of stellar
  mass: the left three panels are for satellite galaxies in different
  halo mass bins, and the most right panel is for central galaxies in all
  haloes  with  masses greater  than  $10^{11.4}  M_{\odot}$.  These
  panels show results  when quenched galaxies are  defined using sSFRs
  which are corrected to total star formation  rates. In each
  panel, the  triangles are the SDSS  data and the colored  lines are
  different  model  predictions  (solid   for  MS,  dashed  for  MS-II
  results)}
\label{fig:ssfr1}
\end{figure}

\begin{figure}
\centering
\includegraphics[scale=0.7]{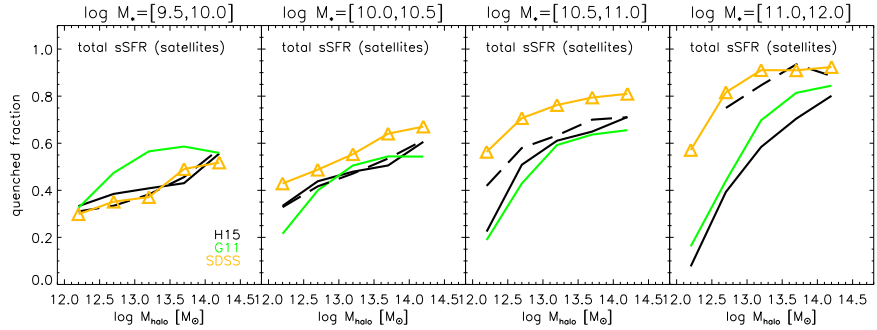}
\caption{As in Fig. \ref{fig:ssfr1}, but now quenched fractions are 
  plotted as a function of halo mass  in different stellar bins.}
\label{fig:ssfr2}
\end{figure}

\subsection{Quenched fraction of satellites and central galaxies}
\label{sect:fq}

In Luo  et al. (2016),  they compared  the quenched fraction  with the
SDSS DR7 data,  based on the model  of Fu et al.  (2013). In agreement
with previous results they also found that there are too many low-mass
quenched satellites  and too few  quenched central galaxies.  Since H15
has fitted the  fraction of passive (quenched) galaxies  as a function
of  stellar  mass  and  redshift,  now we  check  this  fraction  more
carefully in this section.

Following   Luo  et   al.  (2016),   we  define   galaxies  with   $\rm
sSFR<10^{-11}yr^{-1}$ as  quenched galaxies  and select  galaxies with
$\rm  \log_{10}M_*/M_{\odot}>9.5$ from  H15.  The  data points  are  from Luo  et
al.  (2016)  using   the MPA-JHU SDSS DR7 catalogue and Yang et al.(2007) 
group catalog with
$M_{*}>10^{9.5}\rm{M_\odot}$          at         $z<0.04$          and
$M_{*}>10^{9.5}\rm{M_\odot}$ at $z=0.04\sim0.06$  .  
In   Fig.  \ref{fig:ssfr1}  and
\ref{fig:ssfr2}, we  plot the quenched  fraction of galaxies  at fixed
halo  mass  and fixed  stellar  mass,  but  for central  galaxies  and
satellites separately.

By looking at  Fig. \ref{fig:ssfr1} and Fig.  \ref{fig:ssfr2}, we find
that compared to  G13 model, the H15 model is  able to reproduce the
red fraction  of central galaxies (right  panel in Fig.\ref{fig:ssfr1})
and  that  of  low-mass  satellite  galaxies  with  $\rm
\log_{10}M_*/M_{\odot}<10$ (left panel  in Fig.\ref{fig:ssfr2}). The improvement
is slightly  better from  the MS-II simulation.   Compared to  the G13
model,  H15  increases  the  quenched  fraction  of  central  galaxies
significantly at $\rm \log_{10}M_*/M_{\odot}>10$,  suggesting that increasing the
AGN   feedback  does   play   a  more   role   in  quenching   massive
galaxies. While  at $\rm  \log_{10}M_*/M_{\odot}<10$, the quenching  fractions of
 central galaxies  from H15 and  G13 models  are similar, at  which AGN
feedback is not efficient.

Fig. \ref{fig:ssfr2} is  more interesting.  It shows that  the H15 can
reproduce the  fraction of  red satellites at  the low-mass  end (left
panel),  but at  mediate  stellar mass  of $\rm  \log_{10}M_*/M_{\odot}=[10,11]$
(middle two  panels) the red  fraction is  still lower than  the data.
This prediction is very similar to  that from G13 model from MS.  In
G13 model, the over-prediction of blue satellites at this mass scale
can  be  understood  because  the  central  galaxies  is  also  bluer.
However, the  blue fraction of central  in H15 model now  agrees with
the  data.    We  will   later  see   from  Fig.   \ref{fig:csmf}  and
Fig. \ref{fig:hmsm}  that the stellar mass  of centrals at z=0  in H15
model is  actually lower  than the  data at given  halo mass,  this is
because  they  tune the  model  parameters  to  fit the  stellar  mass
functions at high redshift and the overall red fraction by introducing
strong AGN feedback  in massive galaxies and  a longer reincorporation
time of  gas in low mass  galaxies. In order  to fit the total  SMF at
z=0,  H15 increase  the  star formation  efficiency  in satellites  by
lowering  the  star  formation  threshold and  ignoring  ram  pressure
stripping in  halo mass  less than  $\rm \log_{10}M_{vir}/M_{\odot}<14$.  
But by doing so,  the mediate  stellar-mass galaxies are  now bluer.  

So the right  color of central galaxies and too  blue color of mediate
stellar-mass satellites ($\rm \log_{10}M_*/M_{\odot}=[10,11]$) in H15 model means
that star formation efficiency should be slightly lower for them. This
could be achieved by using a  threshold as function as galaxy mass,  or include 
ram-pressure stripping  of hot  gas in the  mediate stellar-mass
satellites.

And   at   most   massive   stellar  mass   bins   (right   panel   in
Fig.  \ref{fig:ssfr2}), there  is  a significant  discrepancy for  the
quenched fraction of satellites between  MS and MS-II (solid and dashed
lines).  This  resolution  problem  in  SAMs  might be  due  
to  the
inconsistent description of physics for satellites whose subhaloes can
not be resolved in simulation,  or the different merger history 
since the merger 
trees of some massive haloes may be not well resolved at higher redshifts 
in the low resolution simulations. The resolution problem has been 
discussed in detail in Luo et al. (2016).

\begin{figure}
\centering
\includegraphics[scale=0.7]{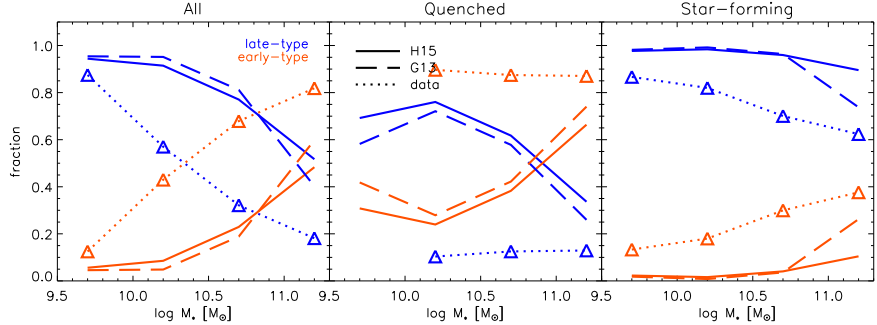}
\caption{The fraction of the central galaxies' morphology as a function 
  of stellar mass. The left-handside panel is for all the central 
  galaxies, middle panel is for the quenched central galaxies, 
  the right-handside panel is for the star-forming central galaxies.
  The red shows the elliptical fraction, black shows the spiral fraction,
  the blue shows the pure-disk fraction only for the Hen15. The triangle 
  is the SDSS data.}
\label{fig:mor1}
\end{figure}

\subsection{The morphology of the central galaxies}
\label{sect:morphology}

It is well known that star  formation activity is close related to the
galaxy morphology (e.g., Kauffmann et  al. 2004), so the morphology is
also a good indicator  of a galaxy to be passive  or star forming.  As
H15 model  now fit the quenched  fraction of the central  galaxies, it
would  be  interested  to  check  if they  are  able  to  predict  the
morphology  of galaxies.  In the  following we  study the  fraction of
galaxies divided into ellipticals and spirals.

In  the  SAMs,  every  galaxy  has   a  disk  plus  a  bulge,  we  use
$B/T=M_{Bulge}/M_{total}$  to define  the  morphology  of the  galaxy.
Normally,galaxies with $B/T \geq 0.7$ are bulge dominated galaxy (i.e.
elliptical galaxy), with $0.7>B/T \geq  0.03$ are normal spirals, with
$B/T<0.03$ are  pure disks  (see Sec.3.8 in  G11).  Similarly,  in the
SDSS  there is  a  photometric parameter  $f_{deV}$  to determine  the
galaxy  morphology. In  SDSS, the  surface luminosity  distribution of
each galaxy  is the linear combination  of the exponential and  the de
Vaucouleurs  profiles, and  $f_{deV}$  is the  coefficient  of the  de
Vaucouleurs term.  Therefore $f_{deV}$ is  most similar to  the $B/T$,
which reflects  the contribution of  the bulge component to  the whole
galaxy.  Bernardi  et al.  (2005)  used  $f_{deV}>0.8$ to  define  the
early-type  (i.e.  elliptical)  galaxies.  Shao  et  al.  (2007)  used
$f_{deV}<0.5$ to define the spiral galaxies. In this work, we use $B/T
\geq 0.7$ as the criterion of  early-type galaxies for SAMs, while use
$f_{deV} \geq  0.7$ at r-band as the  criterion of early-type galaxies  
for our SDSS sample (the  same sample  as Sec.\ref{sect:fq}).  
Correspondingly, the
galaxies  with $B/T<0.7$  or $f_{deV}<0.7$  are selected  as late-type
galaxies.

In  Fig. \ref{fig:mor1},  we plot  the fraction  of central  galaxies'
morphology  as  a  function  of  stellar  mass.  Red  curves  are  for
early-type  (elliptical)  galaxies,  blue  curves  are  for  late-type
(spirals) galaxies.  Solid  and dashed curves are results  for H15 and
G13, triangle dotted curves are results from our SDSS sample.  
The left  panel is  the results  for all  the galaxies,  the
middle and right panels are for quenched and star-forming galaxies.

From the left  panel, we find H15 and G13  have almost same morphology
fractions  and  trend  for   the  whole  sample:  early-type/late-type
galaxies fraction  increases/decreases with stellar mass.  But that is
different  from the  SDSS  data.  There  are  more early-type  central
galaxies and less  late-type central galaxies in the data  than H15 and
G13  models   at  $\rm  \log_{10}M_*/M_{\odot}>10$.  Furthermore,   in  the  data
early-type   is   the   majority   in   central   galaxies   at   $\rm
\log_{10}M_*/M_{\odot}>10.3$  , while  for H15  and G13  model late-types  is the
majority  through all  the stellar  mass. This  indicates that  galaxy
morphology in  the tao models is  quite different with the  SDSS data:
there are more late-type central galaxies in models than in the SDSS.

We further divide the sample into quenched and star-forming galaxies
to  examine the  morphology  fraction. For  quenched central  galaxies
(middle  panel), the  early-type  fraction is  almost  $90\%$ at  $\rm
\log_{10}M_*/M_{\odot}>10$ in SDSS, yet it is  totally different with H15 and G13
models: early-type  dominated only  at $\rm \log_{10}M_*/M_{\odot}>11$.  Even the
fraction trends are different between SDSS and models: the fraction in
SDSS does not  change with stellar mass.  Comparing the  H15 with G13,
we  find their  results almost  similar, and  there are  slightly more
late-type and  less early-type  central galaxies in  H15 than  that in
G13. This implies  there is slightly more  late-type galaxies quenched
in H15 model.  For star-forming  central galaxies, the fraction trends
are almost  similar, whereas in  the models  there are still  more (at
least $10\%$) late-type central galaxies than in the SDSS.

We  can then  conclude  that the  H15 model  has  correct fraction  of
quenched central galaxies,  but the quenching mechanism in  H15 may be
not  correct.  It  seems that  H15 model  quenches too  many late-type
central galaxies in stellar mass range at $\rm \log_{10}M_*/M_{\odot}=[10,11]$. In the
H15 model the quenching is mainly by increasing the efficiency of 
AGN feedback, not related with the increasing  
formation of a bulge. We think  that a more efficient quenching should
be related to  the mass growth in  the bulge and will  discuss this in
the last session.

\begin{figure}
\centering
\includegraphics[scale=0.9]{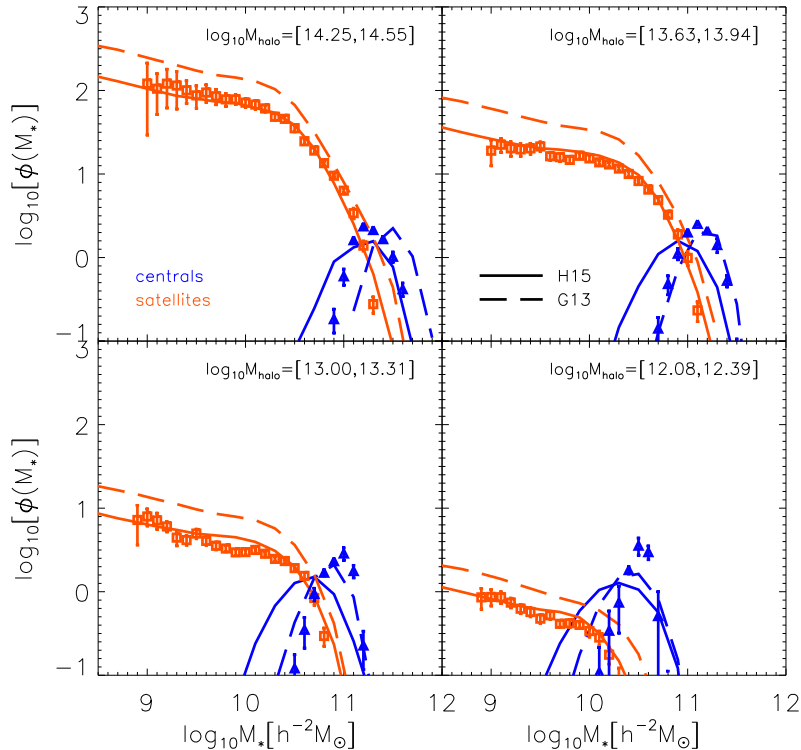}
\caption{CSMFs in different halo mass bins.The solid curves are from H15 
and dashed curves are from G13. The data points are from the 
Yang et al. 2012. The red/blue lines are for satellites/central galaxies.} 
\label{fig:csmf}
\end{figure}

\subsection{The conditional stellar mass functions}
\label{sect:csmf}

The conditional stellar mass function  (CSMF, firstly proposed by Yang
et al. 2003) describes the average number of galaxies as a function of
stellar mass  in dark matter haloes  of given mass. The  CSMF 
which closely relates to correlation functions, provides
additional  constraints  on  galaxy  formation and  evolution,  as  in
principle the formation of dark  matter haloes depend on their mass,
so their  galaxy population  should be different.  Kang et  al. (2012)
have shown that the CSMFs in G13 model is inconsistent with the data
by producing more stellar mass  in satellite galaxies. Since H15 model
provide better  fits to the  observed SMFs at different  redshifts and
the correct fraction  of quenched central galaxies, it  is deserved to
check their model predictions on the CSMFs.

In Fig. \ref{fig:csmf}, we show the CSMFs in different halo mass bins,
and separate the contributions by  central (blue curves) and satellite
(red curves) galaxies. The solid and  dashed lines are for H15 and G13
models, and the data  points are from the Yang et  al. (2012) which is
measured from their constructed group catalog using the SDSS DR7.

We find that, the CSMFs of satellite  galaxies in the H15 model are in
good agreement  with the SDSS  data through  the whole halo  mass, and
they are lower than  the results of G13 about at least  0.2 dex at low
stellar mass ends.   But for central galaxies, except  for the highest
halo  mass bins  $\rm \log_{10}M_{halo}/M_{\odot}=[14.25,14.55]$, G13  model is  more
consistent with the SDSS data than  the H15 model. The stellar mass of
central in H15 model is systematically  lower than the data by 0.2-0.3
dex.  This is related to the high  AGN feedback in H15 model, which is
enhanced  by 1  dex  than G13  model  at  $z=0$, see  Fig.  S4 in  the
supplementary material  of H15. This  fast quenching leads to  a lower
stellar mass in H15 model.

However, the  fact that the  quenched fraction of central  galaxies in
H15 is consistent  with SDSS data, suggests  that increasing quenching
of central galaxies in the model  is necessary. This indicates that we
need  to  quench more  central  galaxies  to  match the  observed  red
fraction,  but only  after the  central galaxy  has the  right amount of stellar
mass.  It is still not clear how  to achieve the two facts at the same
time, we will touch this in the discussion part.

\begin{figure}
\centering
\includegraphics[scale=0.9]{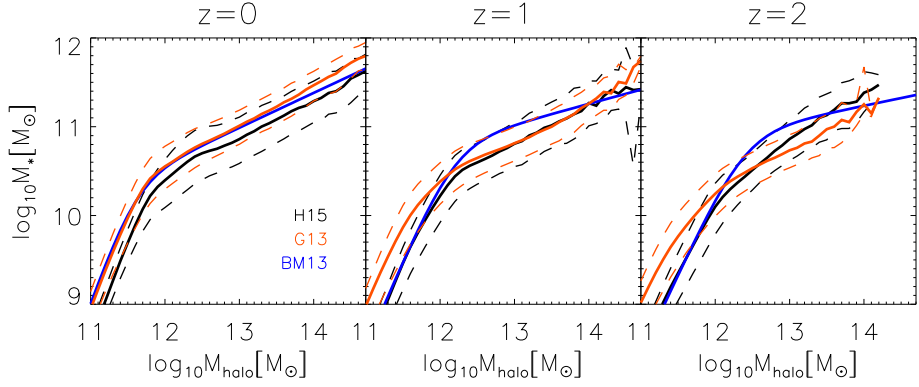}
\caption{The SMHM relation at $z=0,1,2$ for central galaxies. The black 
	lines are the median vaule from the H15, the red lines are the 
	median vaule from the G13, and the dashed lines represent the 68 
	percentile scatter of the models. The blue lines are from the 
	Moster et al. 2013.} 
\label{fig:hmsm}
\end{figure}

\subsection{The stellar mass-halo mass relation}
\label{sect:smhm}

In  previous section  we have  found at  given halo  mass the  central
galaxies in  H15 model are have  lower stellar mass than  the data. In
this section we further comparing  the models prediction of H15, G13
to the data by showing the stellar mass-halo mass (SMHM) relation. The
SMHM relation is a very  basic and important relation for constraining
galaxy  formation physics,  as  shown  by many  others  (e.g., Guo  et
al.  2010;  Kang  et  al.  2012)  that  once  this  relation  is  well
determined, it  would be easily to  fit the stellar mass  function and
galaxy clustering.

We plot the SMHM relations  for central galaxies at different redshift
in Fig. \ref{fig:hmsm},  black lines are from H15, red  lines are from
G13, blue  lines are  from the abundance  matching results  (Moster et
al. 2013).  It is found that at $z=0$ (left panel), at fixed halo mass
stellar mass of  H15 is lower than the abundance  matching results and
H13. At high  redshifts (middle and right panels)  For massive haloes,
the stellar  mass from  H15 and  G13 models  are similar,  except at
$z=2$  the H15  model predicts  a  higher stellar  mass. For  low-mass
haloes, the stellar mass from H15 is also lower than G13 model. This
is consistent with  what H15 did in their model.  H15 used MCMC method
to  find the  best parameters  to fit  the SMFs  at high-z  where they
predicted slightly more massive galaxies at $z=2$ (see their Fig.2) so
their  stellar mass  in massive  haloes at  $z=2$ is  higher. But  for
low-mass haloes,  they adopted  a longer  gas reincorporation  time to
best fit the  faint end of SMFs,  so their stellar mass  is lower than
G13 model.

In the model of  H15, the AGN feedback at $z=0$  is stronger than that
of G13,  so they are able  to produce more red  central galaxies, as
stated before, to fit the red fraction  in the data. This is why their
stellar mass in massive haloes is  lower than both G13 model and the
abundance matching of Moster et al. (2013).

\begin{figure}
\centering
\includegraphics[scale=1.0]{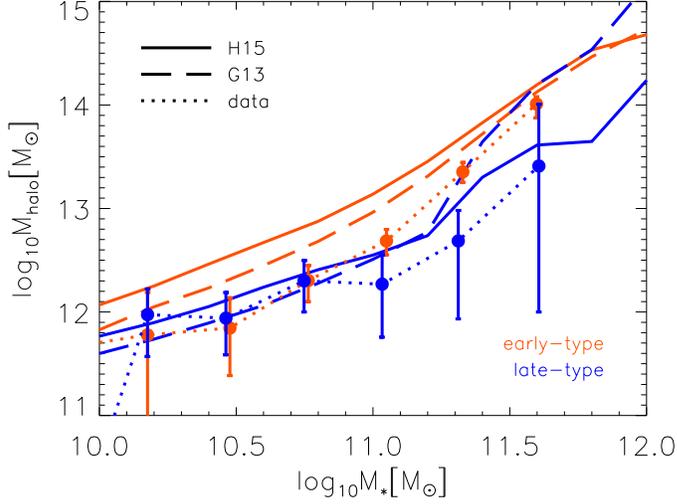}
\caption{The SMHM relation at $z=0$ for central galaxies, but separately 
         for early-type (red) and late-type (blue) galaxies. The data 
  	 points with err bar are form Mandelbaum et al. 2006} 
\label{fig:smhm}
\end{figure}

At  the  SMHM relation  of  Moster  et  al.  (2013) is  obtained  from
abundance  matching,  it  is  not   the  observed  halo  mass,  so  in
Fig. \ref{fig:smhm} we  plot SMHM relation using the  observed data by
Mandelbaum et  al. (2006) where  the halo  mass is measured  from weak
lensing results from  SDSS, which is more reliable.  We separately the
central galaxies  sample into early-type  and late-type, and  show the
results at  $z=0$. Here, early-type is  defined as $B/T \geq  0.7$ for
the models.

It is easy to see from Fig.\ref{fig:smhm} that at given stellar mass, the
early-type galaxies from both G13 and  H15 models are higher than the
data.  We note  that there might be some systematic  difference in the
stellar mass  obtained in the  models and  observations. So we  do not
focus  on   the  absolute  halo   mass  or  stellar  mass   from  this
relation. What is more interesting is difference in the SMHM relations
for the  early-type and  late-type galaxies.  Fig.\ref{fig:smhm} shows
that in the  data (points with error bar) the  halo mass of early-type
and  late-type  galaxies are  very  similar  for $\rm  \log_{10}M_*/M_{\odot}  <
11$.  For more  massive galaxies,  at given  stellar mass  the early-type
galaxies  have higher  halo mass.  But note  that the  error bar  of
late-type galaxies  is very large  at the high  mass end due  to their
lower number density.  However, it is seen that in  both the G13 and
H15  models, the  halo mass  of early-type  galaxies is  higher than
late-type galaxies at  given stellar mass, inconsistent with  the data at
$\rm \log_{10}M_*/M_{\odot}<11$. The  discrepancy is slightly larger  in the H15
model.  This indicates that the  AGN feedback is either over-estimated
in  bulge  dominated galaxies  or  under-estimated  in disk  dominated
galaxies.  As  both G13 and H15  model have too many  disk dominated
galaxies, lowering their  AGN feedback will lead to  more stellar mass
growth and  will ruin the agreement  with the observed SMFs.   We thus
conclude that it is more likely the AGN feedback in the both models is
too strong, and being more stronger in the H15 model.
  
\section{Conclusions and Discussion}

In this  paper, we compare the predictions from the  latest two
versions of the  L-Galaxies model, H15 and G13 to the  SDSS data.   
The H15 model  has introduced  a few
modifications to the  G13 model and is tuned to better  match the
evolution of stellar mass function and the fraction of red galaxies as
a function of stellar mass.  By comparing the model predictions to the
data in more detail, we have obtained the following results:
 
\noindent(I) We examine the quenched  fraction of central galaxies and
satellites   separately.   It  is   found   that,   compared  to   the
over-prediction of blue  central galaxies in G13, the  H15 model can
reproduce the red fraction of central galaxies. The over-prediction of
low-mass red satellites are also solved in H15 model to match the 
data.  But  the  quenched  fraction  of  satellite  galaxies  at  $\rm
\log_{10}M_*/M_{\odot}=[10,11]$ in  both G13  and H15 models  is still
lower than the SDSS data.

\noindent(II) We check the morphology of the central galaxies. We find
there are too many late-type  galaxies and too few early-type galaxies
in both G13  and H15 models compared to the  SDSS data. In addition,
too     many     late-type     galaxies     in     H15     at     $\rm
\log_{10}M_*/M_{\odot}=[10.0,11.0]$ are quenched.

\noindent(III) We  examine the  stellar mass  function of  central and
satellite  galaxies in  different halo  mass bins,  and find  that H15
produces  better  match  to  the observed  stellar  mass  function  of
satellites than  G13 model,  but the  match to  the stellar  mass of
central is worse  than G13. The stellar mass of  central galaxies is
lower by about 0.2 dex than the data suggests.

\noindent(IV) We  compare the stellar  mass-halo mass relation  to the
data, and find  that G13 fits better the  observed stellar mass-halo
mass relation,  and at given  halo mass,  the stellar mass  of central
galaxies in H15 is lower than  the data, especially for the early-type
galaxies.

H15 have clearly  shown that a long reincorporation  time of supernova
ejected gas is helpful to suppress the star formation rate in low-mass
galaxies, thus providing  a better match to the  stellar mass function
at high redshifts.  To solve  the over-red of these low-mass satellite
galaxies ($\rm  \log_{10}M_*/M_{\odot}<10$) they  use a  lower threshold
gas density for  star formation and ignore  the ram-pressure stripping
in low-mass  haloes. However, we found  that such a extension  of star
formation in satellites produce slightly  more blue satellites at $\rm
10<\log_{10}M_*/M_{\odot}<11$. This could be solved  by using a density threshold
as function  as galaxy  stellar mass,  not a  constant in  current H15
model.

By introducing  a AGN feedback  parameter which higher than  the G13
one  at low  redshift, H15  model can  reproduce the  red fraction  of
central galaxies. However, the stellar  mass of those quenched central
galaxies is  now slightly lower (by  about 0.2dex) than the  data. Our
results also show that in the G13  and H15 models there are too many
disk dominated galaxies, and especially too many red disk galaxies. To
solve these  two problems simultaneously,  we need some  mechanisms to
increase the  growth of the  bulge component in central  galaxies, and
the feedback (whatever AGN or others) from the bulge formation is used
to quench the galaxy. In this way,  the stellar mass of central can be
increased to  match the  data, and  we can  have more  bulge dominated
quenched galaxies.

In L-Galaxies, the bulge formation  are through three channels: major,
minor mergers  and disk  instability. For  major mergers,  two galaxies
merger with  mass ratio ($m_{\rm  sat}/m_{\rm cen}$) larger  than 0.3,
then all  the star  in both  two galaxies and  the newly  formed stars
during the merger are assumed to  form a new bulge. For minor mergers,
only  the stars  in  satellite are  added into  the  bulge of  central
galaxy.   Normally,  during major  merger  the  strong starburst  will
consume  almost  all  the  cold  gas  and  suppress  the  further  star
formation.  In L-Galaxies  major mergers  are better  modelled due  to
their rapid  dynamical friction time, so  we do not think  there is no
more space to  increase the major merger rate. On  the contrary, there
is some  freedom to increase  the minor merger  rate, as shown  by van
Daalen et  al. (2016) that a  shorter merger time is  really needed to
match the galaxy clustering using the H15 model.

Disk instability is another important channel  to form a bulge. Due to
this dynamic instability, the unstable disk will transfer some stellar
mass transferred into  bulge to make the disk stable  again. Also some
new stars  can form  during the instability.   G11 has  pointed disk
instability is  a major  way to  form bulge  in the  intermediate mass
galaxies like  the Milky Way.  This is also  the mass range  where the
bulge grow insufficiently  in H15. So increasing  the disk instability
might be the possible solution.

Recently,  Tonini et  al. (2016)  present a  comprehensive theoretical
prescription of  growth of disks  and bulges in SAM.  They distinguish
the   bulges   into   two  populations:   merger-driven   bulges   and
instability-driven   bulges.   Their   model   can   reproduce   some
observations,   such   as   the  mass-size   relation,   Faber-Jackson
relation. They found that  the merger-driven ellipticals are dominated
in  both   low-mass  and  high-mass   ends  of  stellar   mass,  while
instability-driven bulges dominate the  intermediate mass range. These
also indicate  that increasing  the disk  instability is  the suitable
solution to improve  the quenching and morphology  of central galaxies
in  SAMs. We  expect  future solutions  will  reconcile the  conflicts
between stellar mass functions, colour and morphology distributions in
SAMs.

\section{Acknowledgement}

We thank Qi Guo for helpful discussion and Lei Wang for providing the 
SDSS DR7 catalog. The Millennium Simulation data
bases  used in  this paper  and the  web application  providing online
access  to them  were constructed  as part  of the  activities of  the
German  Astrophysical   Virtual  Observatory  (GAVO).  This   work  is
supported  by   the  National   basic  research  programme   of  China
(2015CB857003,    2013CB834900),     NSF    of     Jiangsu    Province
(No. BK20140050), the  NSFC (No. 11333008) and  the Strategic Priority
Research Program 'The Emergence of  Cosmological Structure' of the CAS
(No. XDB09010403).

\end{document}